\documentclass[manuscript]{acmart}
\setcopyright{none}

\usepackage{graphicx}
\graphicspath{ {./images/} }
\begin{document}

\title{Inclusive by Design: Developing Barrier-Free Authentication for Blind and Low\newline Vision Users through the ALIAS Project}
\date{April 2025}

\author{Clara Toussaint}
\email{clara.toussaint@univ-poitiers.fr}
\affiliation{%
  \institution{CeRCA, Université de Poitiers, Université François-Rabelais de Tours, CNRS}
  \city{Poitiers}
  \country{France}
}
\author{Benjamin Château}
\email{benjamin.chateau@cnrs.fr}
\affiliation{%
  \institution{CeRCA, Université de Poitiers, Université François-Rabelais de Tours, CNRS}
  \city{Poitiers}
  \country{France}
}

\author{Emilie Bonnefoy}
\email{emilie.bonnefoy@opensezam.com}
\affiliation{%
  \institution{OpenSezam}
  \city{Paris}
  \country{France}
}

\author{Pierre-Guillaume Gourio-Jewell}
\email{pierre-guillaume.gourio-jewell@opensezam.com}
\affiliation{%
  \institution{OpenSezam}
  \city{Paris}
  \country{France}
}

\author{Nicolas Louveton}
\email{nicolas.louveton@univ-poitiers.fr}
\affiliation{%
  \institution{CeRCA, Université de Poitiers, Université François-Rabelais de Tours, CNRS}
  \city{Poitiers}
  \country{France}
}

\renewcommand{\shortauthors}{Toussaint, et al.}

\begin{abstract}

Authentication is the cornerstone of information security in our daily lives. However, disabled users such as Blind and Low-Vision (BLV) ones are left behind in digital services due to the lack of accessibility. According to the World Health Organization, 36 million people are blind worldwide. It is estimated that there will be 115 million by 2050, due to the ageing of the population. Yet accessing digital services has become increasingly essential. At the same time, cyber threats targeting individuals have also increased strongly in the last few years. The ALIAS project addresses the need for accessible digital authentication solutions for BLV users facing challenges with digital technology. Security systems can inhibit access for these individuals as they become more complex. This project aims to create a barrier-free authentication system based on cognitive ergonomics and user experience (UX) design methods specifically for BLV users.  This paper presents an overview of current research in this area. We also identify research gaps, and finally, we present our project's methodology and approach. First, we will build a knowledge base on the digital practices and cognitive models of BLV users during authentication. This information will support the development of prototypes, which will be tested and refined through two iterations before finalizing the operational version.

\end{abstract}

\ccsdesc[500]{Human-centered computing~Accessibility design and evaluation methods}
\ccsdesc[500]{Security and privacy~Usability in security and privacy}
\ccsdesc[300]{Security and privacy~Biometrics}
\ccsdesc[300]{Security and privacy~Multi-factor authentication}
\ccsdesc[100]{Security and privacy~Privacy protections}
\ccsdesc[500]{Human-centered computing~User centered design}

\keywords{Accessibility, cybersecurity, blind, low vision, authentication, user-centered design}

\maketitle

\section{Introduction}
    The web and mobile applications have become the primary channels for delivering content, services, and products to individuals and professionals. However, as malicious activities continue to rise, digital security has become an imperative. Beyond the often technical approach to cybersecurity \cite{pollini2022leveraging}, considering human factors remains a critical component in risk prevention and a significant source of vulnerability \cite{shabut2016cyber}. Authentication is a cornerstone of digital security: it must ensure secure access, data confidentiality, and the identification of all system actors.
    
    Yet the growing need for security often results in more complex access to services. This dilemma is clearly articulated by Dourish et al. (p.3) \cite{dourish2004security}: "Security systems typically seek to introduce barriers to action (such as passwords or other authentication mechanisms), while HMI (Human Machine Interaction) designers attempt to remove those barriers." For example, systems often require users to enter a code received by email or to solve a CAPTCHA \footnote{Completely Automated Public Turing test to tell Computers and Humans Apar} to reinforce security. However, such barriers can be insurmountable for some individuals, especially people with visual impairments. In France, more than one million people are affected by this disability \footnote{\url{https://drees.solidarites-sante.gouv.fr/publications/etudes-et-resultats/les-personnes-ayant-un-handicap-visuel-les-apports-de-lenquete}}, or nearly 15 out of every 1,000 inhabitants.

    The ALIAS (Authentication Liable for Inclusivity and Simplicity) project was launched to address the accessibility challenges in digital security. According to the Association Valentin Haüy (AVH), which supports individuals with visual impairments, reveals that these people are increasingly connected. AVH states that less than 15\% of blind people can read Braille in France\footnote{\url{https://informations.handicap.fr/a--2732.php}}, and against all intuition, some even use a tactile smartphone, both to communicate and to overcome their daily difficulties. It is therefore essential to understand their uses and involve them in the design of inclusive security solutions. Our goal is to improve our understanding of how individuals with vision impairments use digital technology, an area that still requires further research for better coverage of challenges and potential solutions. We specifically aim to explore how the lack of accessibility and usability affects these users' perceptions of risk. This knowledge is crucial for developing innovative authentication methods that are both safe and inclusive.

\section{Authentication usability}
\label{authentication-usability}
    Authentication is vital to online services, as it verifies the user's identity and safeguards their data from unauthorised access. Authentication is not part of the user's mental model of the task they want to perform (e.g., making a purchase) and, therefore, they will perceive this step as "a barrier" \cite{dourish2004security}.
    
   Authentication can be defined as follows: "The user identifies himself by sending \emph{x} to the system; the system authenticates their identity by calculating \emph{F(x)} and verifying that it is equal to the stored value \emph{y}\textquotesingle{} \cite{lamport1981password}. The most common method is the username-password pair. However, the growth of the Internet has led to malicious behavior, pushing services to enforce more complex password requirements.
    
    Passwords have to be difficult to "guess", which makes the authentication process increasingly burdensome in digital activities. Yet, in its list of usability criteria, Nielsen \cite{nielsen1994usability} encourages interface designers to rely on recognition (using cues to initiate recall) rather than recall (reconstructing information from scratch).
    
    Depending on the level of expertise, sensitivity, and trust of the user \cite{sheehan2002toward,usable2004search}, different password management strategies will be observed (see, for example, the table on the vigilance zone, password management \cite{pollini2022leveraging}) which either go in the direction of a weakening security (use familiar words, customize a base password, keep a paper or electronic list of all passwords \cite{briotto2018understanding}), or in that of advanced management (e.g., password manager).
    
    Weak passwords are particularly vulnerable to hacking algorithms (e.g., brute force), which try all possible combinations \cite{anderson2021security}. This is how CAPTCHAs emerged to counter these methods.
    
    In contrast, biometric solutions such as fingerprint scanning and face recognition stand out for their usability and security. They have the advantage of being "invisible"; they do not require a physical object (a card) or prior knowledge of the user (password) \cite{aleksandr2018multi}. However, they are not always desirable: they imply user confidence in managing personal data, reliability of recognition, and, ultimately, invisibility itself can be a usability problem (e.g., unintentional unlocking of the smartphone by touching the sensor).
    
    From now on, implementing multi-factor authentication (MFA) is becoming increasingly widespread, primarily due to the simplicity of smartphone integration. This method is more secure, but it should be noted that it is also more complex for the user, as it adds authentication steps before allowing access to the service.

\section{Accessibility and authentication}
\label{accessibility-and-authentication}
    Adding authentication factors is an effective solution for protecting users from cyberattacks. However, some authentication methods can present significant challenges for individuals with conditions affecting their vision, hearing, cognition, or motor capacities, like CAPTCHAs \cite{moreno2014captcha}. That is why the challenge should be shifted towards the designers of these systems, to make them more accessible, as recommended by the ISO 9241-171: 2008 standard, to ensure systems are straightforward to use, and thus usable by a wide variety of people with differing abilities. Furthermore, the \emph{Web Accessibility Initiative (WAI)} promotes web accessibility, establishing standards like WCAG\footnote{https://www.w3.org/WAI/standards-guidelines/wcag/} in the USA and RG2A\footnote{https://accessibilite.numerique.gouv.fr} in France to ensure the web is usable by all.
    
    We focus on people who are blind or have low vision (hereinafter BLV), as user interfaces generally communicate information visually, particularly during the authentication process. For this reason, BLV users may have to use assistive tools such as screen readers, magnifiers, voice control, or braille terminals for their digital use. It should be noted that such tools do not always preserve privacy, depending on the authentication context.
    
    This is especially true on the mobile platform, mainly due to the ubiquity of touchscreens \cite{d2016wireless}. Exploring a touchscreen interface for a BLV user can be achieved with the help of assistive tools, which can create a lack of comfort due to potential observers nearby who could easily hear or see what the user is doing. Typing a password on a smartphone is also considered one of the most challenging tasks for a BLV \cite{barbosa2016unipass}. And can cause discomfort when entered in public \cite{ahmed2016addressing} or exposed to shoulder surfing \cite{kumar2007reducing}.
    
    For these reasons, most BLV users do not protect their mobile devices with a password \cite{azenkot2012passchords,dosono2015m}. Even if they consider authentication to be essential or very important (96\% on a questionnaire on 325 people) \cite{briotto2018understanding}.
    
    However, BLV people tend to lower their vigilance concerning their security and digital privacy when surrounded by people close to them; they adopt a transparent approach with their data \cite{akter2020uncomfortable}.
    
    Former research has demonstrated that, among authentication methods, iris scans and patterns drawn on the screen are the least accessible, whilst fingerprinting is the most accessible and secure \cite{briotto2018understanding}. And although PINs are ubiquitous on mobile, they are perceived as uncomfortable \cite{azenkot2012passchords} because they impose a significant slowdown in user activity \cite{wolf2018empirical}. BLV people also perceive them as the least secure method because they are easy to guess and more vulnerable to shoulder surfing \cite{briotto2018understanding}.

    The authors of this literature review \cite{andrew2020review} note that additional devices can be used to increase the security and confidentiality of authentication methods, such as braille passwords \cite{alnfiai2019braillepassword} or digital magnifying glasses \cite{stearns2018design}. They emphasize the importance of integrating these devices into the daily tools of BLV users \cite{shinohara2018tenets}.

\section {Gap in research}
Our reviews of the literature indicate that authentication methods generally exhibit low usability, even for typical users. This limitation is exacerbated for individuals with disabilities, particularly BLV users, due to a lack of accessibility. Poor usability and low awareness of cybersecurity risks can lead to increased dependence on others and weakened security when using digital services.
    
    \emph{Future research should focus on two main areas:}
    \begin{itemize}
        \item First, understanding how information security is perceived within the BLV population and how this perception impacts their daily use of digital services. This requires gathering knowledge about the digital equipment and usage of BLV users. Given the limited literature in this domain, this will necessitate conducting user studies with BLV individuals.
    \end{itemize}
    \begin{itemize}
        \item Second, analyzing how advanced authentication technologies could help design more accessible authentication pathways. While accessibility design standards (e.g., ISO/IEC 30071-1:2019) offer practical guidance, they may fall short when applied to breakthrough technologies. Then it would be valuable to rely on a model grounded in cognitive ergonomics. However, existing research on BLV people’s mental model mainly focuses on spatial cognition (e.g., \cite{guerreiro2020virtual}) and rarely addresses Human Machine Interaction (HMI). Developing a cognitive model of the digital authentication activity among BLV users could help support more inclusive innovations in cybersecurity.
    \end{itemize}

\section{ALIAS project framework}
    The ALIAS project aims to bridge the gap in cybersecurity for BLV individuals. It follows an iterative process based on a user-centered design approach. Indeed, UX involves pragmatic qualities (task efficiency and satisfaction) and hedonic qualities (user emotions and preferences). ISO 9241-210:2019 highlights three key factors for assessing UX: the system, the user, and the context (c.f., Fig. 1.). User-Centered Design \emph{(UCD)} is the best approach to enhance these qualities, using iterative design and humanities methodologies to consider the user's perspective. UCD must involve a diverse user base, including those with disabilities.

     \begin{figure}[H]
        \centering
        \includegraphics[width=0.3\linewidth]{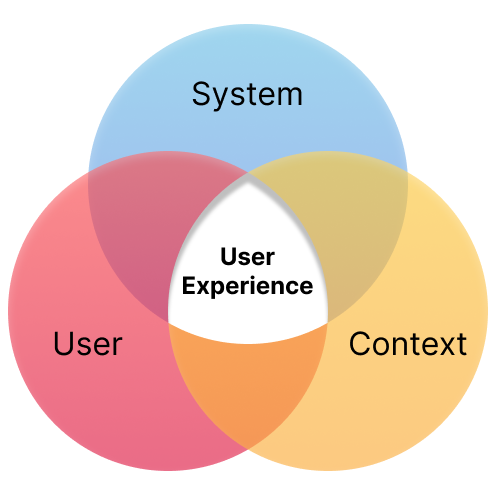}
        \caption{UX key factors as defined by ISO 9241-210:2019. "UX is a consequence of a user’s internal state (predispositions, expectations,
    needs, motivation, mood, etc.), the characteristics of the designed system (e.g., complexity, purpose, usability, functionality, etc.) and
    the context (or the environment) within which the interaction occurs (e.g., organisational/social setting, meaningfulness of the activity,
    voluntariness of use, etc.)" \cite{hassenzahl2006user}}
        \label{fig:diagram1}
    \end{figure}

    We follow a three-stage approach to adapt these user-centered design requirements to our research project (c.f., Fig. 2). The first stage will study digital service usage and the perceived risks and challenges related to traditional authentication systems. To achieve this, a questionnaire survey made on an accessible platform called Qualtrics (target sample > 300 BLV individuals, as \cite{briotto2018understanding}) will be implemented to describe the digital environment of BLV users, particularly in terms of equipment, usage, and reliance on human or technological assistance. The questionnaire will be complemented by focus groups (target sample > 30 BLV individuals) to gather feedback on using authentication services and further explain the questionnaire results. Each focus group consists of two workshops. The first is an experience-sharing session on the use of numeric and cybersecurity, and the second involves a mental simulation of different types of authentication. In order to guarantee the accessibility of focus groups, consent will initially be requested without recording, and subsequently with recording. Visual support, such as images and videos, will not be used. The second workshop will encourage oral description of authentication steps with different models (password, CAPTCHA, magic link, biometrics). The results from this first stage will be used to build a cognitive model of the authentication activity among BLV users, guiding the design team in developing an Alpha version of a more inclusive and acceptable authentication solution.
    
    The second stage will again place BLV users at the center of the process. The goal will be to provide suggestions for improvement, both for developing an Alpha version of the solution and for refining the cognitive model developed in the first stage. These suggestions will be derived from the results of a user test conducted on usable website mock-ups. The test material will be designed to evaluate main authentication systems (username, password, e-mail/SMS code, etc.) as well as a preliminary version of an accessible authentication solution, developed based on Stage 1 findings. Each participant (n=14) will interact with all authentication systems following the Think-Aloud Protocol \cite{ericsson1993} and will then evaluate each system using a questionnaire \cite{venkatesh2008technology}. The test materials will be validated beforehand through a focus group (n=4).

    The third stage will be identical to the second, aiming to validate the directions taken in the Beta version and provide recommendations for the final version. This time, the tests will focus on two existing websites (one private service and one public service). For each website, users will evaluate an accessible version using conventional authentication, as well as an adapted version integrating the Alpha version of the solution. The experimental protocol and sample sizes will be identical to those used in the second stage.

    \begin{figure}[H]
        \centering
        \includegraphics[height=0.22\linewidth]{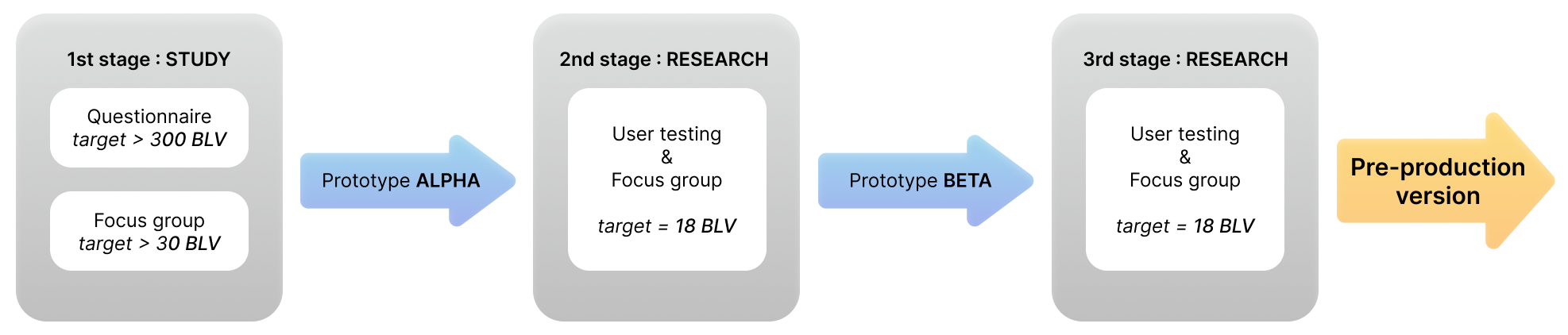}
        \caption{ALIAS Project Stages}
        \label{fig:diagram2}
    \end{figure}

\section{Conclusion}

    Ensuring that our digital world is accessible and secure for all users is increasingly critical in contemporary society. In this context, authentication is a fundamental component of information security. However, it often presents significant barriers for many individuals, particularly those who are blind or visually impaired. The ALIAS project seeks to address this issue using cognitive ergonomics and a user-centered design approach to enhance our knowledge of how risk perception, technology design, and disability interact to lead to unsafe behavior and digital exclusion. Ultimately, the ALIAS project aspires to establish a framework that integrates scientific inquiry, technological advancement, and ethical considerations about technology and security design for all.

\begin{acks}
     This work was supported by a French government grant managed by the French National Research Agency under the France 2030 program, reference ANR-22-PTCC0001.
\end{acks}

\bibliographystyle{ACM-Reference-Format}
\bibliography{bibliography.bib}

\end{document}